\documentclass[aps,prc,reprint,superscriptaddress,floatfix,longbibliography]{revtex4-2} 

\usepackage{xspace}
\newcommand{\pentatrap}{\textsc{Pentatrap}\xspace}

\usepackage{dcolumn}
\usepackage{tikz}
\usepackage{pgfplots}
\usepgfplotslibrary{groupplots}
\usetikzlibrary{arrows.meta}%
\usepgfplotslibrary{dateplot}
\usetikzlibrary{shapes,arrows, patterns}%
\usepgfplotslibrary{colormaps}%
\pgfplotsset{compat=newest}
\usetikzlibrary{pgfplots.groupplots,
               pgfplots.fillbetween}
\usetikzlibrary{decorations.pathreplacing}
\usetikzlibrary{decorations.pathreplacing}
\usepgfplotslibrary{dateplot}
\usepackage{romannum}

\usepackage{hyperref, url}

\def\equationautorefname~#1\null{Eq.~(#1)\null}

\begin{document}


\title{Atomic mass determination of uranium-238}

\author{Kathrin Kromer}
\email[]{Corresponding author.\\ kromer@mpi-hd.mpg.de}
\affiliation{Max-Planck-Institut für Kernphysik, 69117 Heidelberg, Germany}
\author{Chunhai Lyu}
\affiliation{Max-Planck-Institut für Kernphysik, 69117 Heidelberg, Germany}
\author{Jacek Biero\'{n}}
\affiliation{Institute of Theoretical Physics, Jagiellonian University, 30-348 Krak{\'o}w, Poland}
\author{Menno Door}
\affiliation{Max-Planck-Institut für Kernphysik, 69117 Heidelberg, Germany}
\author{Lucia Enzmann}
\affiliation{Max-Planck-Institut für Kernphysik, 69117 Heidelberg, Germany}
\affiliation{Ruprecht-Karls-Universität Heidelberg, 69117 Heidelberg, Germany}

\author{Pavel Filianin}
\affiliation{Max-Planck-Institut für Kernphysik, 69117 Heidelberg, Germany}
\author{Gediminas Gaigalas}
\affiliation{Institute of Theoretical Physics and Astronomy, 10222 Vilnius, Lithuania}
\author{Zoltán Harman}
\affiliation{Max-Planck-Institut für Kernphysik, 69117 Heidelberg, Germany}

\author{Jost Herkenhoff}
\affiliation{Max-Planck-Institut für Kernphysik, 69117 Heidelberg, Germany}
\author{Wenjia Huang}
\affiliation{Advanced Energy Science and Technology Guangdong Laboratory, Huizhou 516007, China}
\author{Christoph H. Keitel}
\affiliation{Max-Planck-Institut für Kernphysik, 69117 Heidelberg, Germany}
\author{Sergey Eliseev}
\affiliation{Max-Planck-Institut für Kernphysik, 69117 Heidelberg, Germany}
\author{Klaus Blaum}
\affiliation{Max-Planck-Institut für Kernphysik, 69117 Heidelberg, Germany}

\date{\today}

\begin{abstract} 

The atomic mass of uranium-238 has been determined to be $238.050\,787\,618(15)$\,u, improving the literature uncertainty by two orders of magnitude. It is obtained from a measurement of the mass ratio of $^{238}$U$^{47+}$ and $^{132}$Xe$^{26+}$ ions with an uncertainty of $3.5\times 10^{-12}$. The measurement was carried out with the Penning-trap mass spectrometer \textsc{Pentatrap} and was accompanied by a calculation of the binding energies $E_{\text{U}}$ and $E_{\text{Xe}}$ of the 47 and 26 missing electrons of the two highly charged ions, respectively. These binding energies were determined using an \textit{ab initio} multiconfiguration Dirac–Hartree–Fock (MCDHF) method to be $E_{\text{U}} = 39\,927(10)$\,eV and $E_{\text{Xe}} = 8\,971.2(21)$\,eV. The new mass value will serve as a reference for high-precision mass measurements in the heavy mass region of the nuclear chart up to transuranium nuclides.

\end{abstract}


\maketitle


Understanding the nuclear structure of heavy and superheavy elements provides clues about the mechanisms involved in synthesizing them and the reasons for their finite lifetimes~\cite{sobiczewskiDescriptionStructureProperties2007, hamiltonSearchSuperheavyNuclei2013}. Facilities measuring the masses of actinides and transactindes contribute to our understanding of the nuclear structure by examining binding energies and derivative values such as nucleon pairing strengths, two-nucleon separation energies, and shell gap parameters~\cite{audiEvaluationAtomicMasses2001}. This experimental data benchmarks nuclear models which are essential for predicting properties of nuclides not accessible through experiments. It is imperative to test these models in regions where experimental data is becoming available, such as the region of the $N=152$ subshell~\cite{blockRecentProgressExperiments2022}, to predict the next "doubly magic" nuclei after $^{208}$Pb or the center and extent of the "island of stability"~\cite{povesNuclearShellModel2001,giulianiColloquiumSuperheavyElements2019}.

For high-precision mass measurements, Penning-trap mass spectrometry (PTMS) has nowadays become one of the leading methods of choice. PTMS now routinely achieves relative mass uncertainties in the range of $10^{-11}$ on stable or long-lived species~\cite{finkDeuteronProtonMass2020, sasidharanPenningTrapMassMeasurement2023, filianinDirectValueDetermination2021} and in the range of $10^{-9}$ on radionuclides~\cite{dworschakPenningTrapMass2010,hukkanenOddoddNeutronrichRhodium2023,kartheinDirectDecayenergyMeasurement2019}. Ideally, carbon-12 is used as the reference mass, since the unified atomic mass unit u is defined as 1/12 of the mass of carbon-12 in its ground state. However, in reality it is often favorable to measure against a reference nuclide similar in mass to the nuclide of interest, so that many systematic uncertainties can be minimized. It is therefore desirable to have a network of nuclides with well-known masses that covers a wide mass range. The most precisely measured nuclides are sometimes referred to as the ``mass backbone"~\cite{m.wangAme2012AtomicMass2012, guenautExtendingMassBackbone2005}. This mass backbone and other known masses are evaluated in form of the Atomic Mass Evaluation (AME)~\cite{wangAME2020Atomic2021}, which considers all kinds of connections, from inertial to energy measurements.

The heavy mass region beyond uranium relies heavily on measurements relating them to the mass of a few uranium isotopes, whose masses are currently known to a precision of at best $5 \times 10^{-9}$~\cite{wangAME2020Atomic2021}. In order to further research into nuclear structure, it is imperative to surpass this precision to benchmark advanced nuclear models. To overcome the limitation set by the reference, we have performed an ultra-precise mass measurement on uranium-238, thereby providing a significantly improved reference mass in the heavy mass region above lead, which contributes to the AME mass backbone.

In addition to serving as a reliable mass reference, an improved atomic mass value of uranium-238 is also needed for the planned investigation of the magnetic moment, and with it the $g$-factor of the bound-electron of hydrogenlike uranium at the experiment \textsc{Alphatrap}~\cite{sturmALPHATRAPExperiment2019, morgnerStringentTestQED2023}. Electron $g$-factors of heavy, highly charged ions provide stringent tests of bound-state quantum electrodynamics (QED) in strong fields as the size of the QED contribution to the $g$-factor increases with the proton number $Z$~\cite{beierG_jFactorBound2000}. However, the precision of a $g$-factor measurement is directly limited by the knowledge of the mass of the ion of interest. In order to achieve a determination of the electron $g$ factor with a precision on the level of $10^{-9}$, the mass of the ion has to be known to the same precision. 

In this letter, we will combine a Penning-trap mass ratio measurement and \textit{ab initio} multi-configuration Dirac-Hartree-Fock (MCDHF) binding-energy calculations to determine the atomic mass of uranium-238.


A determination of the mass of an ion $m$ with charge $q$ in a Penning trap is based on the measurement of the free cyclotron frequency $\nu_c = qB/(2\pi m)$ of the ion in a static homogeneous magnetic field $B$. In order to confine the particle's motion in all three dimensions, a Penning trap is composed of an electrostatic quadrupolar field in addition to the magnetic field. The combination of both fields forces the ion on a trajectory consisting of three independent eigenmotions (small to large in order of the size of the eigenfrequency): the magnetron motion with frequency $\nu_-$, the axial motion with frequency $\nu_z$, and the modified cyclotron motion with frequency $\nu_+$. In order to obtain the free cyclotron frequency, the relation 
\begin{equation}
    \nu_c^2 = \nu_-^2 +\nu_z^2 + \nu_+^2 
     \label{eq:invariance}
\end{equation}
can be used~\cite{brownPrecisionSpectroscopyCharged1982}. Since the magnetic field is not known precisely enough, one measures the cyclotron frequency of the ion of interest (subscripted ioi) with respect to the cyclotron frequency of a reference ion (subscripted ref) with well-known mass $m_{\text{ref}}$~\cite{wangAME2020Atomic2021}. The measured ratio $R$ of the cyclotron frequencies is just proportional to the ratio of the ions' masses, since the magnetic field cancels to first order: 
\begin{equation}
R= \frac{\nu_{c,\text{ioi}}}{ \nu_{c,\text{ref}}} = \frac{m_{\text{ref}}}{m_{\text{ioi}}}\frac{q_{\text{ioi}}}{q_{\text{ref}}}\text{ .}
\label{eq:cyclo_ratio}
\end{equation}

Usually, systematic effects increase with a larger mass difference, however, most systematic effects stemming from various trap imperfections and $B$-field inhomogeneities are minimized when using a similar charge-to-mass ($q/m$) ratio of the ion of interest and the reference ion. For this reason, mass measurements at \pentatrap are carried out on a broad range of ion masses and charge states with the flexibility of choosing any reference ion that is most suited for each specific measurement~\cite{reppPENTATRAPNovelCryogenic2012,rouxTrapDesignPENTATRAP2012}. For the determination of the absolute mass of uranium-238, the near $q/m$ doublet $^{238}$U$^{47+}$ and $^{132}$Xe$^{26+}$ was chosen with a difference in $q/m$ of $3.24\times 10^{-4}\,e/\text{u}$.

\begin{figure*}
    \centering
    \includegraphics[width=173mm]{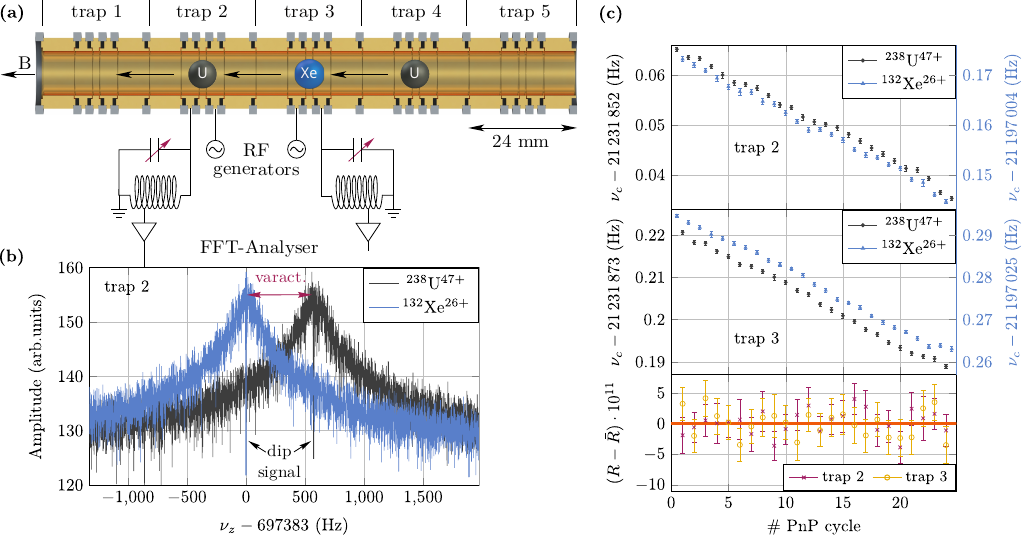} 
    \caption{(a) Sectional view of the Penning-trap tower with three ions in configuration 1. Configuration 2 is indicated by arrows. (b) Axial spectrum of trap 2 showing an overlay of a resonator with dip signal of the Xe ion and of the U ion at their respectively different resonator frequencies varied by the varactor. (c) Exemplary measurement run showing the determined cyclotron frequencies of both ions (y axis of the xenon ion is on the right in blue) and the ratios that can be formed by interpolation in time of the cyclotron frequencies (lower panel). }
    \label{fig:combination_plot}
\end{figure*}

Highly charged ions are delivered to the mass spectrometer via a time-of-flight selective beamline~\cite{kromerHighprecisionMassMeasurement2022} from a Heidelberg compact EBIT~\cite{mickeHeidelbergCompactElectron2018}, equipped with a laser-desorption setup~\cite{schweigerProductionHighlyCharged2019}. A small uranium laser target was used for the uranium ion production and a collimating gas-inlet system introduces xenon gas into the EBIT. The desired charge states of U and Xe were guided through the beamline by electrostatic lenses and a bender, time-of-flight selected by a pulsed operation of a Bradbury-Nielson gate~\cite{bradburyAbsoluteValuesElectron1936} and slowed down by two pulsed drift tubes. For further information on the beamline, see \cite{kromerHighprecisionMassMeasurement2022}. Afterward, the slow ions can be captured inside the Penning trap tower made up of five individual traps. Two of the inner Penning traps are used for frequency measurements, and the other three are used for ion storage. There are in total three ions loaded in alternating sequence [see \autoref{fig:combination_plot}(a)]. This way, the three ions can be moved up or down from configuration 1 to 2 between frequency determinations, effectively swapping the ion species in each of the measurement traps. On the one hand, this double measurement scheme allows for a doubling of measurement statistics since traps two and three are used in parallel to measure a cyclotron frequency ratio each. On the other hand, the two traps have different electric and magnetic field parameters, thus allowing for the cyclotron frequency ratio comparison to ensure a reliable evaluation of systematic shifts.

To determine the frequencies of the eigenmotions of the ions, cryogenic RLC resonators are connected to the axially offset electrodes in each measurement trap [see \autoref{fig:combination_plot}(a)]. The ion interacts with the resonator via the image current induced inside the trap electrodes by the axial motion of the ion. The ion's axial frequency can be brought into resonance with the center frequency of the resonator by tuning the trap depth. Once $\nu_z \approx \nu_{\text{res}}$, the ion's axial motional amplitude will be damped to equilibrium with the thermal Johnson-Nyquist noise of the resonator, effectively cooling the ion's axial motion to around 4\,K. Once the ion is cold, the resonator spectrum will show a ``dip" signal at the ion's axial frequency [see~\autoref{fig:combination_plot}(b)]. This non-destructive detection technique is called Fourier-transform ion-cyclotron-resonance (FT-ICR)~\cite{fengTankCircuitModel1996}. In order to determine the radial frequencies and reduce the radial amplitudes, one can couple each of them to the axial frequency, causing a ``double-dip" from which the frequency of the coupled motion can be deduced~\cite{cornellModeCouplingPenning1990b}. The coupling drive can be induced by a frequency generator connected to a segmented, axially offset electrode.

In each trap, the potential is set to the same trap depth for both ion species, in order to minimize the potential systematic shifts of the measured cyclotron-frequency ratios due to the different ion positions in the traps. However, this results in different axial frequencies of the Xe and U ions due to their different $q/m$ ratios. Variable GaAs capacitors (varactors) were recently implemented into the cryogenic RLC circuits of the traps~\cite{heisseHighPrecisionMeasurementProton2017a} in order to adjust the resonance frequency of the detection circuit with respect to the axial frequency of the ions [see~\autoref{fig:combination_plot}(a,b)].

The largest of the three eigenfrequencies, the modified cyclotron frequency $\nu_+$, is measured phase-sensitively using the pulse-and-phase (PnP) method~\cite{cornellSingleionCyclotronResonance1989}. A PnP sequence consists of an excitation pulse at the modified cyclotron frequency to set the initial phase, then a wait period called phase accumulation time $t_{\text{acc}}$, and finally a radio frequency (rf) $\pi$ pulse at the sideband frequency $\nu_{\text{rf}} = \nu_+ - \nu_z$ to couple the modified cyclotron motion to the axial motion. The $\pi$ pulse transfers not only the energy from the modified cyclotron motion to the axial motion but also its phase information, which can then be read out via the axial resonator, two amplification stages, a subsequent ADC and by applying a Fourier transform. In order to subtract the starting phase and any shifts to the phase by the excitation and readout electronics, a ``short" phase measurement with $t_{\text{acc}}= 0.1$\,s precedes the actual ``long" measurement phase with an accumulation time of $t_{\text{acc}}= 70$ to 100\,s. To reduce the influence of electric field drifts in the trap, the axial frequency is measured during the long PnP phase measurement of the modified cyclotron frequency $\nu_+$ via the dip technique. The magnetron frequency, being the smallest frequency, is only measured once in the beginning of every measurement run via the double dip method. The magnetron frequency of the reference ion is, however, calculated with the help of the magnetron frequency difference measurement, which was performed for the image charge shift measurement campaign (see Supplemental Material Sec.~A). The reason for using the calculated magnetron frequency instead of the measured absolute frequency is, that the cyclotron frequency ratio is more sensitive to the difference of the magnetron frequencies than the absolute frequencies. With measuring the difference instead of using the absolutely measured frequency for the reference ion, we avoid an unnecessarily large uncertainty of the magnetron frequency due to the double dip measurement. \autoref{fig:combination_plot}(c) shows the cyclotron frequencies and ratios in both traps of a measurement run of $\approx12$\,h. The ratios are formed by interpolating the cyclotron frequency of one ion to the point in time of the other ion's cyclotron frequency measurement. With the described measurement scheme, we were able to demonstrate determinations of relative mass ratios with uncertainties of a few $10^{-12}$~\cite{KromerMeta, SchweigerHoDy, HeißeDoor2023}.

The measured cyclotron frequency ratio $\tilde{R} = \nu_c(^{238}\text{U}^{47+}) / \nu_c(^{132}\text{Xe}^{26+})$ is $\tilde{R_2} = 1.001\,644\,000\,787\,9(30)$ and $\tilde{R_3} =1.001\,644\,000\,785\,5(25)$ for trap 2 and trap 3, respectively. This measured ratio was corrected for several systematic effects, see \autoref{tab:systematics}, which are described in detail in the Supplemental Material. 
\renewcommand{\arraystretch}{1.2}
\begin{table}[!htb]
\caption{\label{tab:systematics} The systematic shifts and their uncertainties of the cyclotron frequency ratio determination. A shift $\Delta R$ is given as $\Delta R= \tilde{R} - R$ with $R$ being the unperturbed frequency ratio and $\tilde{R}$ the measured value. The errors of the last three shifts are correlated due to their dependence on the uncertainty of the excitation radii. All values are given in $10^{-12}$. }
\begin{ruledtabular}
\begin{tabular}{ldd}
&\multicolumn{1}{c}{\textrm{Trap 2}}&
\multicolumn{1}{c}{\textrm{Trap 3}}\\
\hline
ICS &-253.1(21) &-257.1(43)\\
Dip lineshape&0.0(11)&0.0(64)\\
Non-linear phase&0.0(6)&0.00(22)\\
Relativistic&0.69(26)&0.5(6)\\
Electrost. anharm. $C_4$ &0.00(23)&0.00(8)\\
Magnetic inhom. $B_2$ &-0.042(12)&0.014(8)\\
\hline
Total systematic&-252.5(25)&-256.6(77)\\ 
\end{tabular}
\end{ruledtabular}
\end{table}
The largest systematic correction comes from the image charge shift (ICS). This effect originates in the interaction between the ion and its image charge on the trap electrodes. The dip lineshape uncertainty originates from the fact that the analytical fit function of the dip spectrum~\cite{fengTankCircuitModel1996} does not describe the spectrum comprehensively. In this case, the fit can yield an axial-frequency value shifted with respect to the true value. The non-linear phase systematic is caused by a non-linear transfer function of the ion's phase during the PnP phase readout. The uncertainty of the difference in motional radii between the uranium and the xenon ion, especially in the excited modified cyclotron motion, adds three correlated systematic effects, namely the relativistic shift, the $C_4$ and the $B_2$ term. The relativistic effect describes the relativistic mass increase of the moving particles. The $C_4$ term approximates the effect that different motional radii have on the trap eigenfrequencies due to electrostatic anharmonicities, in leading order described by the coefficient $C_4$. Similarly, a quadratic inhomogeneity $B_2$ of the magnetic trapping field will also shift the frequencies of the two ions depending on their radii. With these corrections the cyclotron frequency ratios (with the statistical, systematic, and total error in first, second, and third bracket) were determined to be $R_2 = 1.001\,644\,001\,040\,4(30)(25)(38)$ and $R_3 = 1.001\,644\,001\,042\,1(25)(77)(81)$ for trap 2 and trap 3 respectively. The weighted average of the cyclotron frequency ratio is: $R = 1.001\,644\,001\,040\,7(35)$.

The absolute mass of uranium-238 can be determined via the following formula 
\begin{eqnarray}
\label{eq:massU}
m\left(^{238}\text{U}\right)\nonumber= \frac{47 }{26 R} m\left(^{132}\text{Xe}^{26+}\right) +47m_e- E_{\text{U}}/c^2 \text{ with}\\
m\left(^{132}\text{Xe}^{26+}\right)= m\left({}^{132}\text{Xe}\right)-26m_e+E_{\text{Xe}}/c^2 \text{\,.}
\end{eqnarray} 
Here, $R$ is the systematically corrected frequency ratio determined above, $m_e$ signifies the mass of an electron~\cite{zatorskiExtractionElectronMass2017}, $c$ is the speed of light, and $E_{\text{Xe}}=8\,971.2(21)$\,eV is the binding-energy difference between Xe$^{26+}$ and neutral Xe atom determined in our previous work~\cite{kromerHighprecisionMassMeasurement2022,lyuExtremeFieldCalculations2023}. The term $E_{\text{U}}=39.7(16)$\,keV represents the binding-energy difference between U$^{47+}$ and neutral U atom, with the 1.6-keV error bar mainly coming from the large uncertainties in the theoretical ionization potentials (IPs) listed in the NIST atomic database~\cite{kramidaNISTAtomicSpectra2021}. To improve the accuracy of $E_{\text{U}}$, in this work, we will calculate it via the \textit{ab initio} fully relativistic MCDHF and relativistic configuration interaction (RCI) methods~\cite{Grant1970,Desclaux1971,grantRelativisticQuantumTheory2007} implemented in the GRASP2018 code~\cite{froesefischerGRASP2018Fortran952019,jonssonGRASPManualUsers2023,jonssonIntroductionRelativisticTheory2023}. For the sake of computational efficiency, we perform a full calculation for the binding-energy difference $E_{\text{U}}^{6-46}$ between U$^{46+}$ and U$^{6+}$ ions that bear closed-shell ground states, with the IPs of the outermost 6 electrons and the IP of U$^{46+}$ being treated separately. 

In the calculation, the atomic state functions (ASFs) are expanded as linear combinations of configuration state functions (CSFs), which are $jj$-coupled Slater determinants of one-electron orbitals, with appropriate angular symmetry and parity. We first solve the MCDHF equations self-consistently~\cite{Grant1970,Desclaux1971,grantRelativisticQuantumTheory2007} to optimize the radial wave functions of the one-electron orbital under the Dirac--Coulomb Hamiltonian. Then, the RCI method is employed to calculate the contributions from frequency-dependent and frequency-independent transverse photon interactions, the mass shift, and QED effects. Different from previous calculations for Pb$^{4+}$, where the intermediately charged ion Pb$^{22+}$ had been used to bridge the calculations of the correlation energy of the 78 electrons, in this work, we have modified the GRASP2018 code such that we can directly account for the full single and double (SD) electron exchange correlations of the 86 electrons in U$^{6+}$. The results are summarized in Tab.~\Romannum{4} of the Supplemental Materials.
We find that the term $E_{\text{U}}^{6-46}$ is dominated by the single-configuration Dirac--Hartree--Fock binding-energy difference. Such single-configuration calculations give rise to a value of 37\,110.01(8)\,eV, with a contribution of $-0.47(1)$, $-0.02(1)$, and $-0.65(6)$\,eV from the finite nuclear size, the mass shift, and the QED effects, respectively. The Breit interaction and the frequency-dependent transverse-photon interaction together contribute $-16.26$\,eV whose uncertainty will be examined later in the correlation energies. To account for the correlation effects, we systematically expand the size of the CSF basis set by allowing SD excitation of electrons from all the occupied orbitals to the systematically increasing set of correlation orbitals. These correlation orbitals are added and optimized with the layer-by-layer approach~\cite{jonssonGRASPManualUsers2023} up to $n=11$ ($n$ is the principal quantum number), where all orbitals with orbital angular momentum from 0 up to $n-1$ are included. 
By extrapolating to $n=\infty$~\cite{lyuExtremeFieldCalculations2023} we obtain a contribution of 64.7(17)\,eV to $E_{\text{U}}^{6-46}$. The contribution from correlation effects beyond the SD electron excitations are conservatively constrained to be of 6.3(63)\,eV~\cite{kromerHighprecisionMassMeasurement2022,lyuExtremeFieldCalculations2023}. Finally, we arrive at $E_{\text{U}}^{6-46}=37\,164(8)$\,eV, with the uncertainty being dominated by higher-order correlation effects (see Supplemental Material Sec.~E for more details). 

To derive $E_{\text{U}}$, one has to add up the IP of U$^{46+}$ as well as the IPs of the outermost 6 electrons of the uranium atom. For the IP of U$^{46+}$, it is calculated to be 2580.9(1)\,eV based on CSF basis set generated via SD excitations from the $4s$ orbital. For low charged uranium, the first three IPs are known experimentally~\cite{kramidaNISTAtomicSpectra2021}. There is also an experimental value for the IP of U$^{3+}$, but it is around 4\,eV larger than that from a recent theoretical calculation based on the multireference configuration interaction method~\cite{brossMultireferenceConfigurationInteraction2015}. Nevertheless, our calculations are in good agreement with the values presented in ref.~\cite{brossMultireferenceConfigurationInteraction2015}: with CSFs generated via SD excitation of electrons starting from the $6s$ orbital, we arrive at values of 33.12(42), 48.14(42) and 63.15(42) eV for the IPs of U$^{3+}$, U$^{4+}$ and U$^{5+}$, respectively. In total, we obtain $E_{\text{U}}^{0-6}=182.0(20)$\,eV for the total binding energy of the outermost 6 electrons. Thus, the binding-energy difference between neutral uranium and U$^{47+}$ is calculated to be $E_{\text{U}}=39\,927(10)$\,eV which is more than two orders of magnitude more accurate than the NIST value~\cite{kramidaNISTAtomicSpectra2021}. 


By combining the measured cyclotron frequency ratio with the calculated electron binding energies and the literature xenon-132 mass~\cite{wangAME2020Atomic2021}, the atomic mass of uranium-238 was calculated using \autoref{eq:massU} which yields the final value of \hbox{$m\left(^{238}\text{U}\right) = 238.050\,787\,618(15)$\,u}. This value represents an improvement of two orders of magnitude compared to the current literature value of \hbox{$m\left(^{238}\text{U}\right) = 238.050\,786\,9(16)$\,u}~\cite{wangAME2020Atomic2021}. The associated mass excess is correspondingly determined to be $47\,308.367(14)$\,keV. With the reduced mass uncertainty of uranium-238, the atomic mass of uranium-239 which is connected to the 238 mass via a neutron capture process and plutonium-242 connected via a well-known alpha decay energy will be improved as well by a factor of 9 and 1.5, respectively~\cite{wangAME2020Atomic2021}. The mass excess of $^{239}$U is readjusted to be $50\,573.31(17)$\,keV and the one of $^{242}$Pu is $54\,717.3(8)$\,keV.

With the new relative mass precision of $6\times 10^{-11}$ achieved in this work, heavy mass determinations on short-lived nuclei, using the uranium mass as a reference, will not be limited by reference precision for the foreseeable future. A future $g$-factor determination of the bound electron of $^{238}$U$^{91+}$ for tests of bound-state quantum electrodynamics can now be carried out with the same precision as that of the mass~\cite{morgnerStringentTestQED2023}.

\begin{acknowledgments}
This work comprises parts of the Ph.D. thesis work of K.K. to be submitted to Heidelberg University, Germany.
This work is part of and funded by the Max-Planck-Gesellschaft and the DFG (German Research Foundation) – Project-ID 273811115 – SFB 1225 ISOQUANT. The project received funding from the European Research Council (ERC) under the European Union’s Horizon 2020 research and innovation programme under grant agreement number 832848 - FunI. Furthermore, we acknowledge funding and support by the International Max-Planck Research School for Precision Tests of Fundamental Symmetries and the Max Planck, RIKEN, PTB Center for Time, Constants and Fundamental Symmetries.
\end{acknowledgments}

\bibliography{Uranium}
\end{document}



\title{Supplemental material to the paper on the uranium-238 atomic mass determination}

\begin{abstract} 
This Supplemental Material describes in more detail the relevant systematic measurements and estimations as well as the contributions to the theoretical binding energy of the missing electrons of $^{238}$U$^{47+}$ for the determination of the atomic mass of uranium-238. 

\end{abstract}

\maketitle

\section{\label{sec:ICS}Systematic measurement: The image charge shift}
The image charge induced by an ion inside a Penning trap can be used for a non-destructive measurement of the ion's eigenfrequencies and for cooling of the ion's motions. However, the image charge also forms an electric field acting back on the ion and shifting its trap-motion frequencies and hence the free cyclotron frequency. This effect is called image charge shift (ICS)~\cite{vandyckNumberDependencyCompensated1989}. In the case of mass ratio determinations at \pentatrap with large mass differences such as Xe and U, this causes the dominating systematic shift and the leading systematic error since the analytical calculation of the ICS in Penning traps was, so far, only tested experimentally with an uncertainty of 5\,\%~\cite{schuhImageChargeShift2019}. Measurements of the ICS are based on a precise determination of the magnetron frequency $\nu_-$ because this motion has the least sensitivity to magnetic field fluctuations compared to the modified cyclotron frequency and it has a larger ICS compared to the axial frequency. In order to test the ICS to below 5\,\%, it was necessary to measure the magnetron frequency difference between reference and ion of interest with an uncertainty of $<0.25$\,mHz. This was achieved by optimizing the pulse sequence by pulse shaping, making it possible to measure the magnetron frequency phase sensitively by the use of the pulse and phase (PnP) method~\cite{cornellSingleionCyclotronResonance1989}. Only with this phase sensitive method was it possible to determine the magnetron frequency difference $\Delta\nu_-^{\text{exp}}$ between Xe and U to a precision of 40 or 90\,$\mu$Hz, corresponding to an uncertainty of the ICS determination of 0.8\,\% or 1.7\,\%, in trap 2 or 3, respectively.

The PnP scheme for the magnetron motion follows the same principles as for the modified cyclotron motion described in the paper. First, the magnetron motion is excited with a radio-frequency (RF) pulse at frequency $\nu_-$ to set an initial phase. Then, the magnetron motion evolves freely for a time $t_{\text{acc}}$, accumulating its phase. The final magnetron phase is then read out by coupling the magnetron motion to the axial motion, transferring its phase information by an RF $\pi$-pulse at the sideband frequency $\nu_{\text{rf}} = \nu_z + \nu_-$. Finally, the phase can be measured with the axial detection system. In order to subtract any offset phase caused by the RF electronics of the excitation and detection system, this PnP scheme always includes a ``short" reference phase ($t_{\text{acc}} = 0.1$\,s) which is subtracted from a ``long" phase ($t_{\text{acc}} = 80$\,s) to calculate the measurement phase. The experimental magnetron frequency differences between a U and a Xe ion are given in \autoref{tab:ICS} as $\Delta\nu_-^{\text{exp}}$. 
\renewcommand{\arraystretch}{1.2}
\begin{table}[!htb]
\caption{The magnetron frequency difference between a U and a Xe ion is calculated as $\Delta \nu_- = \nu_-(^{238}\text{U}^{47+}) - \nu_-(^{132}\text{Xe}^{26+})$. $\Delta \nu_-^{\text{exp}}$ is the experimentally measured frequency difference, while $\Delta \nu_-^{\text{id}}$ is the calculated one and its uncertainty is due to the systematic shifts except for the ICS. With the resulting difference between these two values due to the image charge shift $\Delta\Delta\nu_{-}^{\text{ICS}}$, an effective trap radius $r_{\text{eff}}$ can be calculated. For the experimental value the uncertainty is given as statistical, systematic, and total uncertainty in the first, second, and third bracket, respectively. Other than that only the total uncertainty is noted. \label{tab:ICS}}
\newcolumntype{d}{D{.}{.}{2.9} }
\begin{ruledtabular}
\begin{tabular}{ldd}
&\multicolumn{1}{c}{\textrm{Trap 2}}&\multicolumn{1}{c}{\textrm{Trap 3}} \\
\hline
$\Delta\nu_-^{\text{exp}}$ (mHz)&-4.83(4)(2)(4)&3.23(4)(8)(9)\\
$\Delta\nu_{-}^{\text{id}}$ (mHz)&-10.212(20)&-2.235(20)\\ 
\hline
$\Delta\Delta\nu_{-}^{\text{ICS}}$ (mHz)&5.38(4)&5.46(9)\\
$r_{\text{eff}}$ (mm)&5.036(14)&5.010(28)\\
\end{tabular}
\end{ruledtabular}
\end{table}

\begin{figure}[b]
    \centering
    \includegraphics[width= 86mm]{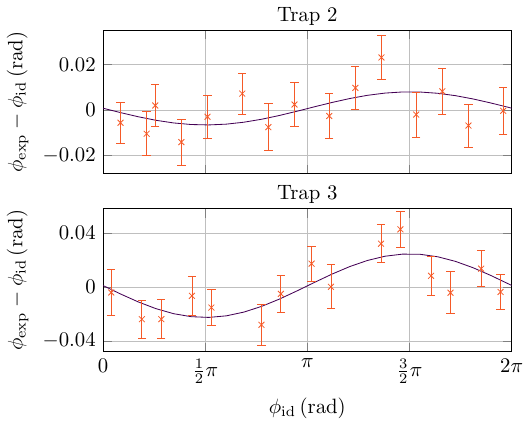}
    \caption{Measurement data (orange) and sine fit (purple) of the nonlinearity measurement of the magnetron phase. The ideal reference phase $\phi_{\text{id}}$ is calculated from the magnetron frequency. The deviation of the measured phase $\phi_{\text{exp}}$ from the ideal phase is plotted on the y axis.\label{fig:nonlinear}}
\end{figure}

The systematic uncertainty of $\Delta\nu_-^{\text{exp}}$ stems from a non-linearity in readout phase originating from the $\pi$-pulse when transferring the phase information from the magnetron to the axial motion. This means that the readout phase is not precisely equal to the phase that the magnetron motion had before coupling but has an additional non-linear (sinusoidal with amplitude $A$) transfer function dependent on the ion's phase, see~\autoref{fig:nonlinear}.
The amplitude of the effect was measured to be for trap 2: $A_2 = 0.0073(29)$\,rad and for trap 3: $A_3 = 0.0236(48)$\,rad. The largest effect on the magnetron frequency by this phase modulation will come from the short reference phase because the long (80\,s) phases fluctuate over time due to voltage instabilities, so the modulation will cause an increase in statistical phase uncertainty but no systematic shift. The reference phase, however, with an accumulation time of only $0.1$\,s, does not fluctuate because it is insensitive to small changes in voltage due to its low frequency resolution. By combining the amplitude with the reference phase difference between the U and Xe ions $\Delta \phi_{\text{short}} \approx 1.1$\,rad in trap 2 and $\Delta \phi_{\text{short}} \approx 2.2$\,rad in trap 3, one can calculate an upper limit of the systematic uncertainty by considering the steepest phase gradient:
\begin{equation}
\Delta \nu_- = \frac{2 \sin(\Delta \phi_{\text{short}}/2)A}{2\pi \Delta t_{\text{acc}}} \text{\,,}
\label{eq:nonlinear}
\end{equation}
with $\Delta t_{\text{acc}}$ representing the difference of the phase accumulation times of the short and the long phase. The uncertainty is calculated to be 20\,$\mu$Hz for trap 2 and 80\,$\mu$Hz for trap 3.

The analysis of the image charge shift follows the principles described in detail in~\cite{schuhImageChargeShift2019}. First, one has to determine the ideal magnetron frequencies of Xe and U without any systematic shifts by: 
\begin{equation}
    \nu_{-}^{\text{id}} = \frac{\nu_c}{2}- \sqrt{\frac{\nu_c^2}{4}- \frac{\nu_z^2}{2}} \text{\,.}
    \label{eq:nu_m_ideal}
\end{equation}
These ideal frequencies $\nu_{-}^{\text{id}}$ are calculated by using the measured frequencies for uranium, taken from one of the measurement runs of the uranium mass determination and then calculating the frequencies for Xe using the values of the magnetic and electrostatic field one can extract from the uranium frequencies. Although the measured frequencies are altered by systematic shifts, the influence of these shifts on the calculation of the magnetron frequency by \autoref{eq:nu_m_ideal} is only on the order of a few $\mu$Hz, so we can still consider this an ideal magnetron frequency difference $\Delta\nu_{-}^{\text{id}}$ for this purpose. The difference of the ideal magnetron frequencies can be found in \autoref{tab:ICS}. This ideal magnetron frequency difference has to be corrected for different frequency shifts of a ``real" Penning trap and the remaining difference to the measured magnetron frequency can then be identified as the image charge shift. Effects that shift the magnetron frequency away from the ideal value include the quadratic magnetic inhomogeneity $B_2$, electrostatic anharmonicities terms like $C_4$ and the tilting angle between magnetic and electrostatic trapping fields as well as an ellipticity of the otherwise harmonic electric field due to e.g. manufacturing tolerances or patch potentials on the trap electrodes. A list of the relevant trap parameters can be found in~\autoref{tab:trapparams}.
\begin{table}[!htb]
\caption{\label{tab:trapparams} Trap parameters relevant in the analysis of the systematic effects. If there are two values given, the first one concerns $^{238}$U$^{47+}$ and the second one $^{132}$Xe$^{26+}$. The magnetic field was calculated using the cyclotron frequency $\nu_c$ from a dip/double dip measurement of $^{132}$Xe$^{26+}$ using the literature value for the atomic mass of Xe~\cite{wangAME2020Atomic2021}. For numbers that were determined or simulated prior to this work, the source is given.}
\begin{ruledtabular}
\begin{tabular}{lllr}
&
\multicolumn{1}{c}{\textrm{Trap 2}}&
\multicolumn{1}{c}{\textrm{Trap 3}}& source\\
\hline
$r_0$\,(mm)&5.000(5)&5.000(5)&\cite{reppPENTATRAPNovelCryogenic2012, rouxTrapDesignPENTATRAP2012}\\
\textit{TR}&0.87987(18)&0.87966(7)&\\
$U_0$\,(V)&-33.9&-15.85&\\
$Q$-factor&$\approx$ 3800&$\approx$ 9400&\\
$\nu_+$\,MHz&$\approx 21.22/21.19$&$\approx21.22/21.19$&\\
$\nu_z$\,kHz&$\approx 698.0/697.4$&$\approx 477.5/477.1$&\\
$\nu_-$\,kHz&$\approx 11.5$&$\approx 5.4$&\\
$B_0$\,(T)&7.002\,147\,55(33)&7.002\,155\,20(33)&\\ 
$B_2$\,$\left(\frac{\text{mT}}{\text{m}^2}\right)$&27.7(1.9)&-4.7(2.1)&\\
$C_4$\,$\left(\frac{1}{\text{mm}^4}\right)$&$0.0(15)\times 10^{-7}$&$0.0(06)\times 10^{-7}$&\cite{rouxTrapDesignPENTATRAP2012}\\
$C_6$\,$\left(\frac{1}{\text{mm}^6}\right)$&$0.0(19)\times 10^{-7}$&$0.0(19)\times 10^{-7}$&\cite{rouxTrapDesignPENTATRAP2012}\\
\end{tabular}
\end{ruledtabular}
\end{table}
The shift of $\Delta \nu_-$ due to $B_2$ scales mainly with the difference in thermal cyclotron radii of the two ion species which for the heavy ions used in this measurement resulted in a negligible shift of $<1\times10^{-4}$\,mHz with respect to the total error of $\Delta\nu_{-}^{\text{id}}$ (see \autoref{tab:ICS})~\cite{ketterFirstorderPerturbativeCalculation2014}. Anharmonicities in the electrostatic field $C_k$, see~\autoref{tab:trapparams}, will not shift the the magnetron frequency difference on a relevant level. The uncertainty on the leading anharmonicity term $C_4$ causes an uncertainty on the magnetron difference calculations of $7 \times 10^{-3}$ and $3 \times 10^{-3}$\,mHz in trap 2 and trap 3, respectively~\cite{ketterFirstorderPerturbativeCalculation2014}.

The impact of an angle $\theta$ between magnetic field and trap axis, as well as an ellipticity $\epsilon$ on the calculated magnetron frequency difference between uranium and xenon has to be determined without knowing a priori the tilt and ellipticity of our traps. The systematic shift of the magnetron frequency difference $\Delta\Delta \nu_-$ can be described by~\cite{rauHighprecisionMeasurementDeuteron2020}
\begin{equation}
\label{eq:angleellips}
    \Delta\Delta \nu_- \approx \Delta \nu_{-} \left(\frac{3}{4} \theta^2 \left(1+\frac{1}{3}\epsilon \cos(2\phi)\right) + \frac{1}{2}\epsilon^2 \right) \text{\,,}
\end{equation}
in which the angle $\phi$ can be set to 0°. Both, the ellipticity and the angle $\theta$ manifest as a mismatch $\Delta\nu_c^{\text{sb,i}}$ between the cyclotron frequency measured using the invariance theorem vs. using the sideband relation($\nu_c = \nu_+ + \nu_-$)~\cite{gabrielseWhySidebandMass2009}:
\begin{equation}
\label{eq:sidebandinvariance}
     \Delta\nu_c^{\text{sb,i}} = \nu_c^{\text{sb}} - \nu_c^{\text{inv}} = \nu_- \left(\frac{9}{4} \theta^2 - \frac{1}{2} \epsilon^2\right)\text{~.}
\end{equation}
While $\Delta\nu_c^{\text{sb,i}}$ was measured during the measurement campaign, this is insufficient to solve for the two unknowns angle $\theta$ and ellipticity $\epsilon$. We have therefore conservatively estimated that the tilting angle of our traps will not exceed $\theta < 0.0262$\,rad $\approx 1.5$\,°. Using the estimated worst-case value of the angle $\theta$, the measured value for $\Delta\nu_c^{\text{sb,i}}$, \autoref{eq:angleellips}, and \autoref{eq:sidebandinvariance}, one can calculate an upper limit of the ellipticity of our traps and the systematic shift to be $\epsilon < 0.055$ and $\Delta\Delta \nu_-< 0.02$\,mHz, respectively. To be sure this angle estimation is large enough one can compare the calculated ellipticity of $\epsilon < 0.055$ to an ellipticity purely due to the electrode's machining tolerances of 5 $\mu$m~\cite{reppPENTATRAPNovelCryogenic2012}, which would give a limit of $\epsilon < 1 \times 10^{-3}$. This value lies comfortably below the one in our estimation. Since these are estimates, no shift was subtracted from the ideal magnetron frequency difference but its error was increased to a value of $0.02$\,mHz.

The differential ICS $\Delta\Delta\nu_{-}^{\text{ICS}}$ can now be calculated as the difference between the experimental and the ideal magnetron difference:
\begin{equation}
    \Delta\Delta\nu_{-}^{\text{ICS}} = \Delta\nu_-^{\text{exp}} - \Delta\nu_-^{\text{id}} 
\end{equation}
The ICS can be analytically calculated in case of an infinitely long cylinder with radius $r_0$. However, the real Penning trap electrodes are not infinitely long and they have slits between them which are necessary to apply the electrostatic trapping field and excitation pulses. The deviation from the ideal case of an infinitely long cylinder can be approximated by calculating an effective trap radius $r_{\text{eff}} = r_0 +\delta$. From the differential ICS in the magnetron mode one can calculate back to these effective trap radii $r_{\text{eff}}$ for each trap, see~\autoref{tab:ICS}, which we then used to determine the ICS of the free cyclotron frequencies during the mass measurements, see~\autoref{tab:systematics}. 
This is done with the following equations, which are the analytical solutions for the case of an infinitely long cylinder but substituting $r_{\text{eff}}$ for $r_0$~\cite{haeffnerPraezisionsmessungMagnetischenMoments2000}:
\begin{equation}
    \Delta\Delta\nu_{-}^{\text{ICS}} = \frac{q_{\text{U}} - q_{\text{Xe}}}{8 \pi^2\epsilon_0 r_{\text{eff}}^3 B_0}
\end{equation}
\begin{equation}
   \frac{ \Delta\nu_{c}}{\nu_{c}} = \left(-\frac{\nu_+}{\nu_c}+\frac{\nu_-}{\nu_c}\right) \frac{m}{4\pi\epsilon_0 r_{\text{eff}}^3 B_0^2} \text{ .}
\end{equation}

\begin{table}[!htb]
\caption{\label{tab:systematics} The systematic shifts and their uncertainties of the cyclotron frequency ratio determination. A shift $\Delta R$ is given as $\Delta R= \tilde{R} - R$ with $R$ being the unperturbed frequency ratio and $\tilde{R}$ the measured value. The errors of the last three shifts are correlated due to their dependence on the uncertainty of the excitation radii. All values are given in $10^{-12}$. \footnote{This table can also be found in the main paper.}}
\begin{ruledtabular}
\begin{tabular}{ldd}
&\multicolumn{1}{c}{\textrm{Trap 2}}&
\multicolumn{1}{c}{\textrm{Trap 3}}\\
\hline
ICS &-253.1(21) &-257.1(43)\\
Dip lineshape&0.0(11)&0.0(64)\\
Non-linear phase&0.0(6)&0.00(22)\\
Relativistic&0.69(26)&0.5(6)\\
Electrost. anharm. $C_4$ &0.00(23)&0.00(8)\\
Magnetic inhom. $B_2$ &-0.042(12)&0.014(8)\\
\hline
Total systematic&-252.5(25)&-256.6(77)\\ 
\end{tabular}
\end{ruledtabular}
\end{table}

\section{Systematic effect: The dip lineshape effect}

The thermalized ion’s axial motion coupled to a cryogenic tank circuit reveals itself in the Johnson-Nyquist-noise frequency spectrum of the tank circuit (further: resonator spectrum) as a dip at the frequency of the ion’s axial motion (further: dip spectrum). A plot of an exemplary dip spectrum in trap 2 can be found in the main article. The analytical fit function of the dip spectrum is described in~\cite{fengTankCircuitModel1996}. The fit of an ideal dip spectrum yields the same, ‘true’, value of the axial frequency regardless of the position of the dip with respect to the resonator spectrum. In practice, the dip spectrum might be subject to shape distortions due to various effects. In this case the ideal fit function does not correctly describe an experimental dip spectrum and hence can yield an axial-frequency value shifted with respect to the true value. Furthermore, $\nu_z$ extracted from the dip-fit might depend on the position of the dip with respect to the resonator center frequency $\nu_{\text{res}}$ and on the fit parameters.

The systematic effects not included in the uncertainty calculated by the covariance matrix of the axial dip-fit can be split into two highly correlated effects. The first effect is the influence of the fit parameters on the determination of $\nu_{\text{res}}$. The second effect comes from the use of the varactor, which we use to shift $\nu_{\text{res}}$ to match with $\nu_z$ of either Xe or U. Both effects, since highly correlated via $\nu_{\text{res}}$, will be summed up in the end. 

In the following two sections, if there are two numbers given, the first always refers to trap 2 and the second to trap 3. 

\subsection{Resonator center frequency uncertainty}
\label{sec:diplineshape1}
The variation of the fit parameters of the resonator spectra in trap 2 and trap 3 yielded a variation of the fitted $\nu_{\text{res}}$ within $\pm0.9$\,Hz and $\pm1$\,Hz, respectively. The size of the effect on the axial frequency by this $\nu_{\text{res}}$ uncertainty can be determined by fitting the dip spectrum using different resonator center frequencies. The fit of the dip spectra in trap 2 and trap 3 yielded, for small detunings, a dependence of the axial frequency on the $\nu_{\text{res}}$ of $0.001\,2(6) / 0.010\,7(7)$\,Hz per 1\,Hz detuning for uranium ions and $0.000\,70(25) / 0.005\,9(8)$\,Hz per 1\,Hz detuning for xenon ions. The effect varies between different traps and ion species due to the different $Q$-factors of the resonators and different dip widths. The resonator $Q$-factors can be found in \autoref{tab:trapparams}. This amounts to a systematic uncertainty on the cyclotron-frequency ratio of $0.9\times 10^{-12} / 5.1\times 10^{-12}$.

\subsection{Axial and resonator center frequency differences}
\label{sec:diplineshape2}
In order to adjust $\nu_{\text{res}}$ to the dip frequency, we alter the capacitance of the resonant tank circuit by employing varactors in the circuit. A change of the varactor voltage not only shifts the resonant frequency of the tank circuit, but can also distort the shape of the resonator spectrum, e.g. by a different noise background at different frequencies. In order to determine the dependence of the axial frequency on the varactor voltage, we carried out a systematic measurement, moving $\nu_{\text{res}}$ around the axial frequency by a few Hz. The resulting uncertainty amounts to an axial shift of 0.000\,15(8) / 0.001\,4(4)\,Hz per Hz resonator shift. The different orders of magnitude of the effect in trap 2 and 3 are due to the difference in $Q$-factor of the resonators in the respective traps, see~\autoref{tab:trapparams}, and by extension the different dip widths. If both ions' axial frequencies are detuned from the center frequency of the resonator by the same amount, the systematic effect cancels out to first order when forming the cyclotron frequency ratio. If the axial frequencies of uranium and xenon relative to their respective resonator center frequencies are different: 
\begin{equation}
    \Delta_{\text{U/Xe}} = \Delta\nu_{\text{U}}  -\Delta \nu_{\text{Xe}} \neq 0 
\end{equation}
with $\Delta \nu_{\text{ion}} = \nu_{\text{res,ion}} -\nu_{z,\text{ion}}$, there will be a systematic uncertainty. The difference $\Delta_{\text{U/Xe}}$ averaged over all measurement runs is $0.83/0.88$\,Hz. By multiplying this with the measured axial shifts one gets a systematic uncertainty of the cyclotron frequency ratio of $2\times 10^{-13} / 1.3 \times 10^{-12}$.

Overall, the conservative uncertainty due to the two dip lineshape effects described in \autoref{sec:diplineshape1} and \autoref{sec:diplineshape2} amounts to $1\times10^{-12}$ in trap 2 and $6\times 10^{-12}$ in trap 3.

\section{Non-linear Phase effect}

As described in~\autoref{sec:ICS}, the $\pi$-pulse during the PnP sequence necessary to transfer the phase information from one mode to another can produce a shift of the ion's read-out phase. This modulation of the original modified cyclotron phase during transfer to the axial motion was measured and fitted with a sine function. The amplitudes for trap 2 and 3 were determined to be $A_2 = 0.002\,9(26)$\,rad and $A_3 = 0.004\,5(44)$\,rad. The long phases of the PnP cycle drift over time by several $2\pi$\,rad due to a slow loss of magnetic field due to the flux creep effect~\cite{andersonTheoryFluxCreep1962}. Therefore, the non-linear phase effect of the long phases presents itself as an increase in statistical uncertainty. However, the short cyclotron phases are stable over time and will have to be considered as a possible source of a systematic effect. \autoref{eq:nonlinear} is also valid for the case of the modified cyclotron phase. The difference of the short phases $\Delta \phi_{\text{short}}$ between the U and the Xe ions are $\Delta \phi_{\text{short}} = 2.26(24)$\,rad in trap 2 and $\Delta \phi_{\text{short}} = 0.71(29)$\,rad in trap 3. Since the systematic effect will be calculated as an uncertainty and not a shift, the sign of the phase difference does not play a role and the absolute values are given. With these values a worst-case systematic uncertainty can be calculated, using \autoref{eq:nonlinear} and $\Delta t_{\text{acc}} \geq 69.95$\,s, resulting in $6\times 10^{-13}$ in trap 2 and $2.2\times 10^{-13}$ in trap 3. The non-linear phase effect depends strongly on the coupling time, pulse shape and frequency and will therefore have to be measured with every new measurement run at \pentatrap. In the future, it is planned to randomize the phase of the excitation pulse which will randomize the effect as well, thereby omitting the systematic shift in favor of a larger statistical error. 

\section{Systematic effects due to different cyclotron radii}
Several systematic shifts can be caused by a difference in the radius of the cyclotron motion between reference and ion of interest during the pulse and phase (PnP) cycle. The difference in excitation radius occurs if the transfer function of the excitation pulse to the trap electrodes is frequency-dependent. Since $^{132}$Xe$^{26+}$ and $^{238}$U$^{47+}$ have a significant difference in the modified cyclotron frequency of $\approx 35\,850$\,Hz, we cannot exclude a different amplitude of the excitation pulse at the position of the ion. In order to determine the size of this effect, we measured the difference in radius of the excited cyclotron motion $\Delta \rho_{+\text{,exc}}$ between the two different ion species in each trap. In order to do so, we shifted the voltages of the correction electrodes away from the optimized voltage to create a strongly anharmonic potential and then measured the difference in axial frequency between the ion with an excited cyclotron motion versus with a cooled cyclotron motion. This difference in axial frequency contains the information of the excited cyclotron radius, which can then be compared between the two ion species~\cite{ketterFirstorderPerturbativeCalculation2014}. With this method we determined the ion radius after the excitation pulse used during the measurement for trap 2/trap 3 respectively: $13.6(4)/ 19.6(5)$\,$\mu$m for $^{132}$Xe$^{26+}$ and $13.4(4)/ 19.4(5)$\,$\mu$m for $^{238}$U$^{47+}$. Thus, the cyclotron radii of U and Xe might differ on a level of 1-2 percent.

For calculating the excitation radii and the following systematic shifts, a few necessary trap parameters are summarized in \autoref{tab:trapparams}.

\subsection{The relativistic shift}
Every moving particle will have a relativistic mass increase of its rest mass $m_0$ of:
\begin{equation}
    m = \frac{m_0}{\sqrt{1-(v/c)^2}} \text{, }
\end{equation}
with $v$ representing the particle's velocity and $c$ the speed of light. During a PnP cycle the radius of the modified cyclotron motion $\rho_+$ is excited to define the starting phase. As described above, this excitation radius $\rho_{+,\text{exc}}$ might not be identical for both ions, and the relativistic mass increase might, therefore, not cancel out completely when forming the cyclotron frequency ratio. The remaining systematic ratio shift arising from the relativistic mass increase can, due to the strong frequency hierarchy, be approximated, using just the modified cyclotron frequency $\nu_+$: 
\begin{equation}
    \frac{\Delta \nu_c}{\nu_c} = 1- \frac{m_0}{m} \approx - \frac{v_+^2}{2c^2} = - \frac{\omega_+^2\rho_{+,\text{exc}}^2}{2c^2} = - \frac{2\pi^2\nu_+^2\rho_{+,\text{exc}}^2}{c^2} \text{.}
\end{equation}
This approximation has been shown to be valid in~\cite{ketterClassicalCalculationRelativistic2014}. The resulting systematic frequency ratio shifts were calculated to be: $6.9(26)\times 10^{-13}$ and $5(6) \times 10^{-13}$ for trap 2 and 3, respectively.

\subsection{Electrostatic imperfections: $C_4$}
\pentatrap uses the typical cylindrical trap geometry with 5 electrodes with different lengths but the same radius. The lengths of the trap electrodes are chosen such that the potential best resembles a harmonic potential in the trap center when applying a certain compensation voltage $V_c$ to the correction electrodes~\cite{rouxTrapDesignPENTATRAP2012}. The traps are mirror symmetric with respect to the plane of the central ring electrode. This allows to describe the potential $\Phi$ in the trap center as a series of Legendre polynomials $P_k$ with purely even orders~\cite{gabrielseOpenendcapPenningTraps1989}: 
\begin{equation}
\label{Legendre}
    \Phi(z, \rho) = V_0 \sum_{k = 1}^{\infty}C_k(z^2+ \rho^2)^{k/2}P_k \left(\frac{z}{\sqrt{z^2 + \rho^2}}\right)
\end{equation}
in cylindrical coordinates $z, \rho$. The coefficients $C_k$ depend on the geometrical structure and the applied voltages. Before each measurement campaign $C_4$ is optimized via the tuning ratio $TR = U_c/U_0$ to be as small as possible. The optimized values for the uranium campaign can be found in~\autoref{tab:trapparams}. The remaining uncertainty of $C_4$ causes an uncertainty of the cyclotron frequency ratio of 2.3 and $0.8\times 10^{-13}$ in trap 2 and 3, respectively. For the individual shifts of each of the eigenmodes, see~\cite{ketterFirstorderPerturbativeCalculation2014}. 

Other systematic shifts dependent on the different cyclotron radii such as higher order electrostatic imperfections e.g. $C_6$ lie far below $10^{-13}$ and can therefore be set to zero.

\subsection{Magnetic inhomogeneity $B_2$}
The magnetic field of the \pentatrap magnet is not perfectly homogeneous. The remaining quadratic magnetic inhomogeneity $B_2$ at the center of the traps causes small shifts of the eigenmotions. The first order of these shifts is described in~\cite{ketterFirstorderPerturbativeCalculation2014}. With this, the $B_2$ shift can be calculated to be $-4.2(1.2) \times 10^{-14}$ for trap 2 and $1.4(8)\times 10^{-14}$ for trap 3.

\section{Contributions to the binding energy difference between U$^{6+}$ and U$^{46+}$}

For U$^{6+}$ with a ground state of [Hg]$6p^{6}~{}^1S_0$, the Dirac–Hartree–Fock calculation based on this single-configuration gives rise to a binding energy of $763\,659.36$\,eV for a point-like nuclear charge. However, with a finite-size nucleus, one obtains a correction of $488.16(40)$\,eV under a Fermi model~\cite{dyallGRASPGeneralpurposeRelativistic1989, parpiaRelativisticBasissetCalculations1992}, with the uncertainty coming from the inaccurate nuclear radius~\cite{angeliTableExperimentalNuclear2013}. In the following, all calculations are based on such an extended nuclear potential. First, the corresponding mass shift (MS) is $-1.39(24)$\,eV with an accuracy to the order of $(m_e/M)(\alpha Z)^4m_ec^2$ ($M$ is the mass of the nucleus). 
Then, the transverse photon interaction contributes an energy of $-1\,023.22$\,eV from the frequency-independent terms (Breit interaction), with the frequency-dependent terms adding $27.33$\,eV to the total binding energy. Furthermore, the vacuum polarization (VP) and self-energy (SE) effects are accounted for via a screened-hydrogenlike model~\cite{grantRelativisticQuantumTheory2007} in the GRASP2018 package~\cite{froesefischerGRASP2018Fortran952019, jonssonGRASPManualUsers2023,jonssonIntroductionRelativisticTheory2023}. With values $-231.41$ and $931.63$\,eV, respectively, these QED effects reduce the binding energy of U$^{6+}$ by $700.22$\,eV. To estimate its accuracy, the QED effects of U$^{88+}$ are calculated with the GRASP2018 code as well. With a value of $625.02$\,eV, it is $8.61$-eV larger than the accurate \textit{ab initio} result~\cite{malyshevQEDCalculationGroundstate2014,malyshevInitioCalculationsEnergy2021}. Assuming a similar systematic error, we obtain a QED contribution of $-690.6(96)$\,eV to the total binding energy of U$^{6+}$. Nevertheless, all of the above corrections are significant only for inner-shell electrons such that most of their effects in U$^{6+}$ cancel with the corresponding effects in U$^{46+}$. As a consequence, as shown in the third row of Table~\ref{tab.mcdhf}, the term $E_{\text{U}}^{6-46}$ is dominated by the $37.1$-keV DHF energy difference, with contributions of $-0.47(1)$, $-0.02(1)$, and $-0.65(6)$\,eV from the finite nuclear size (FNS), the MS and the QED effects, respectively. 
\begin{table*}
\caption{Different contributions to the total binding energies of $^{238}$U$^{46+}$ and $^{238}$U$^{6+}$: DHF$_0$ the DHF energy assuming a point-like nuclear charge; FNS, the finite nuclear size effect; MS, the mass shift; Breit, the frequency-independent transverse photon interaction; ${\omega}$TP, the frequency-dependent transverse photon interaction; QED, the QED contribution based on screened-hydrogenic model; SDc, the correlation energies arising from single and double electron excitations; HOc, the systematic effect from all unaccounted correlation effects. The uncertainties of DHF$_0$, Breit and ${\omega}$TP terms are accounted for as a whole in the SDc and HOc terms. The final results are round up to integer values. All entries are shown in units of eV. }
\begin{tabular*}{\textwidth}{@{\extracolsep{\stretch{1}}}*{11}{llrrlrrrlcr}@{}}
  \hline
   ions & ground state &  DHF$_0$ & FNS  & MS   &   Breit & ${\omega}$TP & QED & SDc  & HOc & total \\
   \hline
  U$^{6+}$ & [Hg]$6p^{6}~{}^1S_0$  &   $763\,659.36$   &  $488.16(40)$ & $-1.39(24)$  &   $-1\,023.22$ & $27.33$ & $-690.6(96)$ & $158.3(15)$ & $8.6(86)$ &  $761\,650(20)$\\
  
    U$^{46+}$ & [Kr]$4d^{10}~{}^1S_0$   & $726\,549.35$  & $487.68(40)$ &  $-1.38(24)$ &   $-1\,007.03$ & $27.41$ &   $-689.9(96)$ & $~\,93.57(37)$ &  $2.3(23)$ &  $724\,487(13)$ \\

    $E_{\text{U}}^{6-46}$ &     & $37\,110.01$  & $-0.47(1)\,~$ &  $-0.02(1)$ &   $-16.18$ & $-0.08$ &   $-0.65(6)$ & $~\,64.7(17)$ &  $6.3(63)$ &  $37\,164(8)$ 

\label{tab.mcdhf}
\end{tabular*}
\end{table*}
While the uncertainties of the differential FNS and MS terms are assumed to bear the same relative uncertainties as those in individual ions, the uncertainty of the QED correction is conservatively given as 10\% of the differential QED contribution. For the energies of the DHF and the transverse photon interactions, their values depend on the basis employed in the calculation. Such a basis dependency is resolved after fully taking into account the correlation effect. Therefore, their uncertainties will be accounted for in the correlation energies discussed below. 

To account for the correlation energies, we systematically expand the size of the CSF basis set by allowing single and double (SD) excitation of electrons, in both U$^{46+}$ and U$^{6+}$ ions, from all the occupied orbitals of the ground-state configuration to high-lying correlation orbitals. These correlation orbitals are added and optimized via the layer-by-layer approach~\cite{jonssonGRASPManualUsers2023} up to $n=11$ ($n$ is the principal quantum number), where all orbitals with orbital angular momentum from 0 up to $n-1$ are included. 
At each layer, the increment of the correlation energy decreases exponentially as a function of $n$~\cite{lyuExtremeFieldCalculations2023}. Thus, via extrapolating to $n=\infty$, we obtain a SD correlation energy of $93.57(37)$ and $158.2(15)$\,eV for U$^{46+}$ and U$^{6+}$, respectively. The uncertainties are the differences between extrapolation results based on different number of data points.  

Furthermore, there are correlation effects beyond the SD electron exchanges. These effects are difficult to evaluate, but can be constrained from the theoretical and experimental IPs of low charged ions. To achieve this aim, we calculated the IP of Fr-like Th$^{3+}$, which has a $5f$ valence electron outside of the [Hg]$6p^6$ core (the IP of Fr-like U$^{5+}$ is not experimentally known to high accuracy). Based on the SD excitations from the $5s$ orbital to virtual orbitals up to $n=10$, the calculated IP of Th$^{3+}$ is 0.42\,eV smaller than the experimental value of $28.648(25)$\,eV. Though SD excitations starting from the $4s$ orbital generate 4 million CSFs at $n=8$ for the expansion of the ground state of Th$^{3+}$ with $J=5/2$, the calculated IP is projected to be 0.94\,eV smaller than the experimental values. Further inclusion of core orbitals below the $4s$ orbital in the calculation of Th$^{3+}$ would generate CSF basis sets intractable with the computer cluster we used. 

In order to detect the systematic effects arising from the $1s-3d$ orbitals, the binding-energy difference between U$^{6+}$ and U$^{24+}$, noted as $E_{\text{U}}^{6-24}$, are calculated based on three schemes -- containing CSFs generated via SD excitations starting from the $1s$, $4s$ and $5s$ orbitals, respectively. With an average value of 5\,725.72\,eV, the three results agree with each other within 1.4\,eV, indicating that the SD core--core correlation contributions to $E_{\text{U}}^{6-24}$ bear a significant cross cancellation between the two ions. Nevertheless, to account for the uncertainties arising from higher-order correlation effects, we conservatively assume the corresponding systematic shift decreases linearly from 1.0\,eV for U$^{6+}$ to 0.1\,eV for U$^{24+}$, and then assume a maximum shift of 0.1\,eV for all ions with higher charges~\cite{lyuExtremeFieldCalculations2023}. With this, the total higher-order correlation effect is constrained to be less than 12.65\,eV. To cover this effect, we add a correction of 6.3(63)\,eV to the binding-energy difference and obtain $E_{\text{U}}^{6-46}=37\,164(8)$\,eV.

\bibliography{Uranium}